\documentclass{article}
\usepackage{a4wide}
\usepackage{graphicx}
\usepackage[english]{babel}
\usepackage{graphicx}
\usepackage{wrapfig}
\usepackage{epsfig}
\usepackage{cite}
\usepackage{paralist}
\usepackage{amsmath}
\usepackage{amssymb}
\usepackage{listings}
\usepackage[ruled,vlined,linesnumbered]{algorithm2e}
\usepackage{tikz}

\usepackage[colorlinks=false,
pdfborder={0 0 0}]{hyperref}

\usepackage{subfigure}
\usepackage{framed}

\usepackage{makecell}
\usepackage{stackengine}
\usepackage{tabulary}

\graphicspath{{./images/}}

\newcommand{\figlabel}[1]{\label{fig:#1}}

\newcommand{\figref}[1]{Fig.~\ref{fig:#1}}

\begin{document}
\title{All That Glitters Is Not Gold \\{\Large Towards Process Discovery Techniques with Guarantees}\thanks{Preprint submitted to the International Conference on Advanced Information Systems Engineering, 2021}}

\author{
	 Jan Martijn E. M. van der Werf \footnote{Corresponding author: j.m.e.m.vanderwerf@uu.nl} \\ Utrecht University 
\and Artem Polyvyanyy \\ The University of Melbourne 
\and Bart R. van Wensveen \\ Utrecht University 
\and Matthieu Brinkhuis \\ Utrecht University 
\and Hajo A. Reijers \\ Utrecht University 
}
\date{}
\maketitle
\begin{abstract}
The aim of a process discovery algorithm is to construct from event data a process model that describes the underlying, real-world process well. 
Intuitively, the better the quality of the event data, the better the quality of the model that is discovered.
However, existing process discovery algorithms do not guarantee this relationship. 
We demonstrate this by using a range of quality measures for both event data and discovered process models. 
This paper is a call to the community of IS engineers to complement their process discovery algorithms with properties that relate qualities of their inputs to those of their outputs. 
To this end, we distinguish four incremental stages for the development of such algorithms, along with concrete guidelines for the formulation of relevant properties and experimental validation. 
We will also use these stages to reflect on the state of the art, which shows the need to move forward in our thinking about algorithmic process discovery.
\end{abstract}

\section{Introduction}\label{sec:introduction}
Process mining focuses on the extraction of process-related information from event logs, a collection of sequences of actions, each encoding a historical process execution~\cite{Aalst2016}.
Process discovery is a core area in process mining.
It studies algorithms that, given an event log, construct process models that aim to describe the corresponding true process as closely as possible.
One of the main challenges for process discovery is that the true process is unknown, and has to be inferred from a \emph{sample} observed and recorded in the event log~\cite{Buijs2014}.

An algorithm is a sequence of computational steps that transform a given \emph{input} into some \emph{output}~\cite{Cormen2009}.
Different algorithms exhibit different properties, for example, correctness, finiteness, definiteness, effectiveness, and efficiency.
Such properties allow us to choose an algorithm that fulfills a certain need, such as performing a guaranteed correct computation within the desired time bounds.
A process discovery algorithm transforms a given input event log into an output process model.
We usually expect that a process discovery algorithm is finite (terminates after a finite number of computational steps), definite (each computational step is unambiguous), effective (each computational step can be performed correctly in a finite amount of time), and efficient (the fewer or faster computation steps can be executed the better).
However, process discovery algorithms treat quality as a \emph{goal} rather than a guarantee.
That is, process discovery algorithms are designed to construct a ``good'' process model from the input event log~\cite{Aalst2016}, where the ``goodness'' of the model is not established by the internals of the algorithm, but by external measures, e.g. precision and recall.

In this paper, we recommend refining the process discovery goal.
Our recommendation is triggered by the observation that a process discovery algorithm can construct a good model from an event log yet discover a worse model from another event log of better quality~\cite{PolyvyanyySWCM20}.
We argue that process discovery algorithms should come with guarantees formulated in terms of the relationship between the quality of its inputs and outputs.
The present paper makes these contributions:
\smallskip
\begin{compactitem}
\item 
We propose measures for the quality of event logs, both in the presence and absence of a so-called true process. In the former case, we use standard conformance checking measures, while in the latter case we rely on sampling techniques and measures as studied in statistics.
\item 
We provide empirical evidence that existing process discovery algorithms can construct good models from event logs and, at the same time, produce poor models from better logs; 
\item 
We propose four stages for process discovery algorithms to guarantee the intuitively appealing dependency between the quality of input event logs and the quality of output process models as constructed by the algorithms from the logs.
\end{compactitem}
\smallskip

We believe that a next step in the evalulation of process discovery algorithms is necessary for the field to advance.
Several benchmarks (cf.~\cite{AugustoCDRMMMS19}) have identified process discovery algorithms that ``glitter'', that is, algorithms that produce high-quality models on a limited collection of event logs. 
We argue that such benchmarks should be complemented with formal analyses to provide quality guarantees with the algorithms, extending the current state-of-the-art evaluation with statistical methods to establish a relation between log and model quality.
We invite the process mining community to contribute to the discussion of the maturity of process discovery algorithms.
In addition, we encourage the authors of existing and future discovery techniques to establish the proposed guarantees.

The remainder of the paper is structured as follows.
The next section introduces the intuition why process discovery algorithm need to provide guarantees.
A statistical approach to establish event log quality is introduced in Sec.~\ref{sec:sampling}. 
The proposed four stages of process discovery algorithms are presented in Sec.~\ref{sec:designing}, together with empirical evidence that algorithms do not provide such guarantees yet.
Last, Sec.~\ref{sec:relatedwork} and~\ref{sec:conclusion} are devoted to related work, and conclusions, respectively.

\section{Setting the Stage}\label{sec:thestage}

\subsection{Process Discovery and Conformance Checking}

Process mining projects often start by assuming that some underlying process generates an event log that can be observed, recorded, and used for process discovery. 
We refer to this underlying entity as the \emph{true process}. 
Based on the observed log, process discovery algorithms aim to construct a process model that describes the true process well. 
Formally, given a set of activities $A$, an event log $L$ is defined as a multiset over finite sequences, called traces, over $A$.
A discovery algorithm $\mathrm{disc}$ can be described as a relation $\mathrm{disc} \subseteq \mathcal{L}(A) \times 2^{\mathcal{M}(A)}$, where $\mathcal{L}(A)$ and $\mathcal{M}(A)$ are the universe of all possible logs and models over $A$, respectively. 
Some algorithms, such as the ILP-miner~\cite{WerfDHS09}, are non-deterministic, i.e., applying a process discovery algorithm may yield different results for the same input log.
The true process is, however, often unknown~\cite{Buijs2014}. 
Hence, it can only be approximated.

To measure how well the discovered process models describe the behavior recorded in the event log, different measures have been proposed~\cite{Aalst2018ataed}.
\emph{Precision} is a function $\mathrm{prec}: \mathcal{L}(A) \times \mathcal{M}(A) \rightarrow [0,1]$ that quantifies the fraction of behavior allowed by the model that was actually observed.
\emph{Recall} is a function $\mathrm{rec}: \mathcal{L}(A) \times \mathcal{M}(A) \rightarrow [0,1]$ that quantifies the observed behavior allowed by the model, where a value of one denotes perfect conformance between the log and model.
As shown in~\cite{Syring2019,PolyvyanyySWCM20}, the entropy-based precision and recall measures satisfy the requirements proposed in~\cite{TaxLSFA18,Aalst2018ataed}.

Process discovery algorithms are often designed with a specific quality goal in mind. 
Several algorithms have \emph{rediscoverability} as their goal: if the unknown, true process that generated the event log has specific properties, and the event log satisfies certain criteria, then the algorithm discovers the true process.
For example, the $\alpha$-miner has the rediscoverability property for structured workflow nets, imposing log completeness as criterion~\cite{AalstWM04}.
Similarly, the Inductive Miner~\cite{LeemansFA15} can rediscover process trees under the assumption of activity completeness, i.e., every leaf in the tree should occur at least once in the event log.
Other algorithms take different approaches, e.g., to return a model that scores best on one or more conformance measures (e.g.,~\cite{WerfDHS09,WeijtersR11,MedeirosWA07}).

\subsection{Relating Log Quality and Model Quality}\label{sec:principleIdea}

\begin{figure}[t]
	\centering
	\includegraphics[trim=0 100 0 100, width=.9\textwidth]{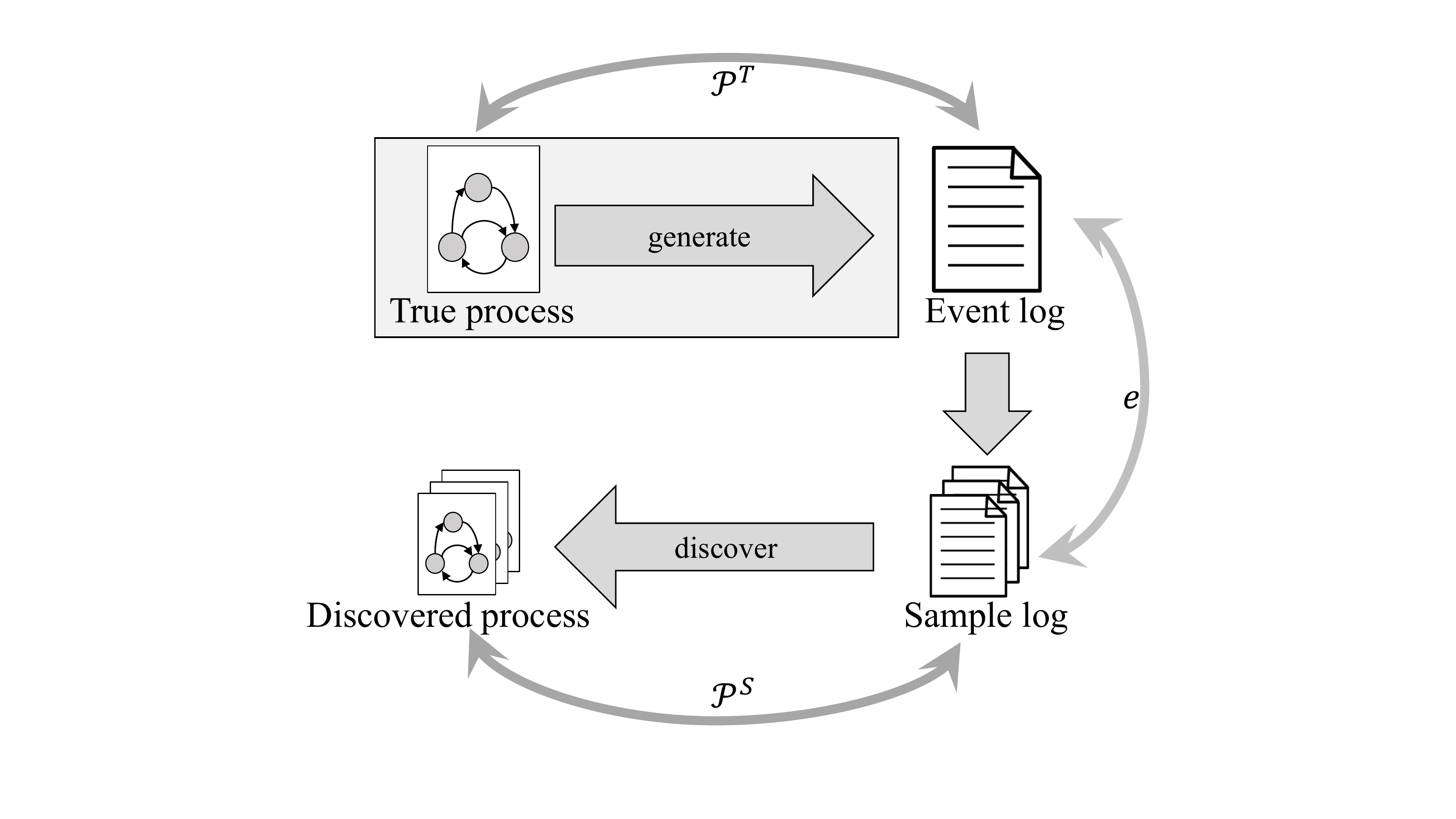}
	\caption{True process generates an event log $L$ with unkown quality $\mathcal{P}^T$. Any sample $S$ drawn from $L$ has some error $e$. Discovering a model from $S$ results in a model with quality $\mathcal{P}^S$.}\label{fig:approach}
\end{figure}

Event logs used as inputs to process discovery algorithms are often assumed to be faithful representations of the true processes.
Let us reflect on the consequences of this assumption. 
Consider~\figref{approach}.
Assume some event log $L$ is a faithful representation of some true process $\mathit{TP}$.
In other words,  $L$ has a high model quality $\mathcal{P}^T$, measured in terms of precision and recall between $L$ and $TP$.
The true process $\mathit{TP}$ is executed continuously, and thus generating a stream of events, from which $L$ is a snapshot~\cite{KnolsW19,Aalst2018ataed}. 
Therefore, $L$ can be seen as a sample from this stream.
Potentially, samples of $L$ can be faithful representations of $\mathit{TP}$ as well. 
Let $S$ be a sample of $L$. 
As it is a sample, the field of statistics provides methods to assess the quality $e$ of the sample with respect to $L$.
And, because $S$ is an event log itself, it can be used to discover some model $M$, which has quality $\mathcal{P}^S$, again measured in terms of precision and recall, but this time between~$S$ and~$M$.
Then, if $S$ is a good representation of log $L$, a process discovery algorithm should construct a model with a quality that approaches $\mathcal{P}^T$. 

Now, draw two samples from $L$, say $S_1$ and $S_2$.
For $S_1$, model $M_1$ is discovered, with model quality $\mathcal{P}^{S_1}$, and for $S_2$ a model $M_2$ is discovered, with model quality $\mathcal{P}^{S_2}$. 
Suppose $S_1$ has a higher sample quality than $S_2$.
In other words, $S_1$ is a better representation for $L$ than $S_2$. 
Intuitively, the quality of $M_1$ should then also be closer to $\mathcal{P}^T$ than the quality of $M_2$.
In other words, if $e(S_1) \geq e(S_2)$ then one should expect that $\mathcal{P}^{S_1} \geq \mathcal{P}^{S_2}$. 
Hence, it is desirable that the process discovery algorithm guarantees that logs of better quality result in models of better quality.

In real-life situations, the true process that generated the event log is unknown.
In most process mining methods (cf.,~\cite{BozkayaGW09,EckLLA2015}) the event log is prepared, and then process discovery techniques are applied to unravel a process model.
An important concern that these methods do not address relates to the internal validity of process mining projects: if the process is repeated on a new observation, i.e., a new event log, to what degree do the results agree between the analyses?
For this property, i.e., test-retest reliability, the guarantees of a process discovery algorithm come into play.
If the different samples are of similar quality, then the constructed models should be of similar quality.
However, current process discovery algorithms do not explicitly claim to provide such guarantees.

\section{Sampling to Measure Log Quality}\label{sec:sampling}
A necessary step in providing guarantees on the results of process discovery algorithms is to establish measures for log quality.
We argue that any event log can be studied as a random sample of traces generated by the true process.
Similar to~\cite{Aalst2018ataed}, the true process can be represented as a set of traces with some trace likelihood function that assigns a probability to each trace.
Consequently, any sample of an event log is again a random sample of the true process, as proposed in~\cite{KnolsW19}.
We consider a sample log $S$ of an event log $L$ to be a subset of the traces observed in the event log, i.e., $S(\sigma) \leq L(\sigma)$, for all traces $\sigma \in L$, and $S(\sigma) = 0$ if $\sigma \not\in L$.
This allows to draw different samples from a given event log, and then compare these samples with the event log to analyze the quality of these samples.
Currently, little is known about the representativeness or quality of random samples in process mining~\cite{KnolsW19,Wensveen2020}.
In the remainder of this section, we propose random sampling techniques to be used in process mining, and provide measures to analyze the quality of a sample with respect to the original event log.

\subsection{Sampling Techniques}
In this section, we propose three probability sampling techniques that can be used to draw a sample from an event log, where each trace in the event log has equal probability of being sampled.
Consequently, samples from these techniques can be used to estimate characteristics of the event log, and, thus, of the true process.

The first technique is \emph{simple random sampling}, where a sample is created by randomly including traces with a predetermined sampling ratio.
The second technique is \emph{stratified sampling}, where the data is divided into unique groups, called strata.
For process discovery, these groups can be formed based on unique traces.
Then, a simple random sample is taken from each group.
In theory, this sampling technique would give more representative samples, because of stratification on unique traces.
However, one has to be careful when applying stratified sampling:
as only a natural number of traces can be added to a sample, a trace can only be added fully or not at all. 
Hence, a problem occurs if a stratum contains fewer traces than there are expected to be sampled.
To solve this, rounding using the half to even rule (cf. IEEE 754) can be used, which rounds halves to the nearest even integer, while still rounding other decimal numbers to the nearest integer.
No literature exists on the topic of using stratified sampling in the area of process discovery~\cite{Wensveen2020}.

An extension of stratified sampling is an approach we call \emph{stratified squared} sampling.
First a stratified sample is drawn, then the number of sampled traces is compared to the number of expected traces based on the sampling ratio.
Suppose the number of expected traces exceeds the number of sampled traces, because of rounding. Then an additional sample is taken by randomly sampling additional traces that have not been included in the sample yet.
This is done by applying a variation of stratified sampling, that samples one case of the most frequent unique trace which has not been sampled yet, until the number of sampled traces equals the number of expected traces or there are no unique traces which are not yet included in the sample left.

\subsection{Sample Quality Measure for Process Mining}
Event logs describe the behavior of a system in terms of traces of events.
As in~\cite{KnolsW19}, we define behavior as the directly-follows relation induced from the event log $L$.
The directly-follows relation $>_L$ is defined on pairs of events $a$ and $b$, such that $a>_L b$ iff the event log $L$ contains a trace in which the two activities $a$ and $b$ occur consecutively.
A first measure to compare a sample to its original event log is existential completeness, i.e., the extent to which all possible directly-follows relations are present.
This results in the first sample quality measure: \emph{coverage}, which is defined by the proportion of unique directly-follows relations present in the sample and the number of unique directly-follows relations in the event log.

Coverage does not take the occurrence frequency of behavior into account.
Different methods exist to measure frequency representativeness.
In statistics, error measures are used to quantify the error between the expected values and the real occurrences.
We propose to adapt these error measures to quantify the error between the behavior observed in a sample, and the expected behavior from the event log based on the sampling ratio.
This results in the following measures for sample quality, where $\mathbf{e}$ denotes the expected behavior, and $\mathbf{s}$ the sampled behavior:

\begin{description}
\item[The Normalised Mean Absolute Error (NMAE)] calculates the normalized absolute deviation (i.e. error) of the number of occurrences of each unique directly-follows relation of the sample from their respective expected frequency:
\[
\textup{NMAE} = \frac{\textup{MAE}}{\textup{avg}\, \mathbf{e}} = \frac{\sum_{i=1}^{n}\left | s_i - e_i \right |}{\sum_{i=1}^{n}e_i}
\]
\item[Normalised Root Mean Square Error (NRMSE)] is similar to NMAE, but uses the root of the squared values, instead of the absolute values, thus penalising large deviations more heavily:
\[
\textup{NRMSE} = \frac{\textup{RMSE}}{\textup{avg}\, \mathbf{e}} = \frac{\sqrt{\frac{1}{n}\sum_{i=1}^{n}(s_i-e_i)^2}}{\frac{1}{n}\sum_{i=1}^{n}e_i}
\]

\item[The Symmetric Mean Absolute Percentage Error (sMAPE)] is a symmetric variation of the NMAE, expressed as a percentage error, with the advantage that the undersampling of behavior is penalised more heavily:
\[
	\textup{sMAPE} = \frac{1}{n}\sum_{i=1}^{n}\frac{\left | e_i - s_i \right |}{e_i + s_i}
\]
\item[The Symmetric Root Mean Square Percentage Error (sRMSPE)] is similar to sMAPE, using the root mean square error instead of the mean absolute error, thus penalising large deviations more heavily:
\[\textup{sRMSPE} = \sqrt{\frac{1}{n}\sum_{i=1}^{n}\left (\frac{e_i - s_i}{e_i + s_i} \right )^2}\]
\end{description}

\smallskip
For a detailed evaluation of the above measures, we refer the reader to~\cite{Wensveen2020}.
These measures assess the behavioral quality of a sample with respect to the event log it is drawn from. 
In other words, these measures provide ways to establish the quality of the input of process discovery algorithms.

\section{Designing Process Discovery Algorithms with Guarantees}\label{sec:designing}
As observed in a study on conformance measures~\cite{PolyvyanyySWCM20}, some process discovery algorithms had a large variability in the quality of the constructed process models. 
In particular, given different samples of a single event log, the same algorithm sometimes provided good results on small samples, while on larger samples, the algorithm discovered worse models.
On further inspection, these algorithms are state-of-the-art, and did not perform any major ``process mining crimes''~\cite{RehseF18}.
In addition, they ``glittered'' in the benchmark study reported in~\cite{AugustoCDRMMMS19}.

We consider this observation as a threat to the application of process mining, in particular for its repeatability and, hence, the reliability of its results.
Suppose for a true process several event logs are captured and analyzed, and the results do not agree, i.e., they differ largely in quality.
Several explanations for this phenomenon are possible. 
A first explanation could be the quality of the input, i.e., the quality of the event logs differed significantly.
However, as the observation highlights, another plausible -- yet undesirable -- explanation lies in the process discovery algorithm itself.
In other words, if the process discovery algorithm does not provide any guarantees on the quality of the resulting models, it is impossible to exclude the algorithm as a root cause. 

Consequently, we advocate process discovery algorithms to provide guarantees on the quality of the produced results. 
To this end, we propose to distinguish four stages during the introduction of a process discovery algorithm:
\smallskip
\begin{compactenum}
	\item The algorithm is well designed;
	\item The algorithm is validated on real-life examples;
	\item The algorithm has an established relationship between the log and model quality;
	\item The algorithm is effective.
\end{compactenum}
\smallskip

\noindent
Though the first two stages are basic, not all algorithms make it to the second stage, as we will illustrate later. 
Arguably, algorithms that are shown not to pass the second stage should not be used in empirical studies. 
The third and fourth stages are entirely novel for process discovery. 
Once the algorithm is shown to be applicable on real-life examples, the authors should study which guarantees their algorithm provides in a controlled setting where the true process is known.
To pass the last stage, the algorithm should provide evidence that in settings where the true process is unknown, the algorithm provides the guarantees stated at stage 3. 
In the remainder of this section, we detail the four stages. 

\vspace{-2mm}
\subsection{Stage 1: The Algorithm is Well Designed}
\vspace{-1mm}
In the first stage, the developers of a process discovery algorithm should properly introduce their algorithm. 
For this, the developers need to provide the following:

\smallskip
\begin{compactitem}
	\item The class of process models the algorithm constructs;
	\item Evidence for meeting the quality goals of the algorithm;
	\item Criteria on the logs, e.g., requirements on the true process that generates the logs;
	\item An initial evaluation on artificial data sets.
\end{compactitem}
\smallskip

\noindent
Most process discovery algorithms satisfy the requirements of this stage.
For example, the ILP-miner~\cite{WerfDHS09} is designed for the class of classical Petri nets with interleaving semantics.
It is proven to always return a Petri net with a perfect recall score. 
It imposes no requirements on the input event logs and is tested on artificial logs.
Also, the $\alpha$-miner~\cite{AalstWM04} algorithm is at least in this stage.
It is designed for well-structured Workflow nets with rediscoverability as a goal. 
It imposes two requirements on an input event log: it should contain all directly-follows relation present in the true process, and the true process should be block-structured.
A similar argument holds for the Inductive Miner~\cite{LeemansFA13}.

\vspace{-2mm}
\subsection{Stage 2: The Algorithm is Validated}
\vspace{-1mm}
Even though an algorithm may be well designed, i.e., it passes stage 1, it is not guaranteed that it works in practice. 
The second stage in introducing the algorithm is, therefore, the validation of the algorithm on a collection of real-life event logs, such as used in the benchmark reported in~\cite{AugustoCDRMMMS19}.
Several algorithms fail to reach this stage. 
For example, the $\alpha$-miner is theoretically a robust algorithm, but the requirements it imposes on the true process are too strong for application in real-life situations. 
Similarly, the ILP-miner is designed from a theoretical point of view and has limitations for practical use, mostly because of its guaranteed recall and runtime performance.
Other algorithms, such as the Inductive Miner, the Declare Miner~\cite{DECLAREMINER} and the Split Miner~\cite{SPLITMINER} have been applied successfully on several real-life event logs, and thus pass this stage.

\begin{figure}[t]
	\centering
	\begin{minipage}[t]{.48\textwidth}
		\null 
		\begin{algorithm}[H]
			{\scriptsize
				\SetInd{0.4em}{1em}
				\caption{Establish Relation}\label{algo:invitro}
				\While{True}{
					$\mathit{TP} \gets$ GenerateModel($\mathcal{M}$, $A$)\;
					\ForEach{$i \in [1..N]$}{
						$L \gets$ GenerateLog($\mathit{TP}$, $T$)\;
						$\mathcal{P}^T \gets$ calcModelQuality($L$, $\mathit{TP}$)\;
						\ForEach{$r \in \mathit{ratios}$}{
							\ForEach{$j \in [1..K]$}{
								$S \gets$ DrawSample($L$, $r$)\;
								$e \gets$ calcSampleQuality($L$, $S$)\;
								\BlankLine
								$M \gets$ DiscoverModel($S$)\;
								$\mathcal{P}^S \gets$ calcModelQuality($S$, $M$)\;
							}
						}
					}
				}
			}
		\end{algorithm}
	\end{minipage}%
	\hspace{1mm}
	\begin{minipage}[t]{.48\textwidth}
		\null
		\begin{algorithm}[H]
			{\scriptsize
				\SetInd{0.4em}{1em}
				\caption{Test Effectiveness}\label{algo:invivo}
				\ForEach{$L \in \mathit{Benchmark}$}{
					\ForEach{$r \in \mathit{ratios}$}{
						\ForEach{$j \in [1..K]$}{
							$S \gets$ DrawSample($L$, $r$)\;
							$e \gets$ calcSampleQuality($L$, $S$)\;
							\BlankLine
							$M \gets$ DiscoverModel($S$)\;
							$\mathcal{P}^S \gets$ calcModelQuality($S$, $M$)\;
						}
					}
				}
			}
		\end{algorithm}
	\end{minipage}
\end{figure}

\subsection{Stage 3: 
	Established Relationship Between Log and Model Quality}
Although passing stage two shows the algorithm's capabilities, this does not provide any guarantees on the quality of the algorithm's output.
As a first step in establishing a relationship between the log and model quality, it needs to be shown to what degree the algorithm satisfies the guarantees as sketched in Fig.~\ref{fig:approach}. 
In other words, the designers need to show that if an event log is a faithful representation of a true process as per measure $\mathcal{P}^T$, then the algorithm should satisfy properties similar to those listed below:

\smallskip
\begin{compactenum}[P1.]
	\item For a sample log $S$ that approaches the perfect quality, the quality $\mathcal{P}^S$ of the discovered model from $S$ approaches $\mathcal{P}^T$;
	\item For two samples $S_1$ and $S_2$, if sample $S_1$ has a higher quality than $S_2$, then the model quality $\mathcal{P}^{S_1}$ is higher than $\mathcal{P}^{S_2}$.
\end{compactenum}
\smallskip

Algorithm designers can choose different strategies to provide evidence for these properties.
The most potent form of evidence is formal proof that the algorithm satisfies these properties for specific instantiations of log and model quality measures. 
In that way, a relationship between an input log quality and the resulting model quality can be established.
We also encourage algorithm designers to define algorithm-specific log quality measures.
If a formal proof is not feasible, instead, statistical evidence of these properties can be provided. 
For this, we propose a controlled experiment as outlined in Algorithm~\ref{algo:invitro}.
Such a controlled experiment follows the approach as shown in Fig.~\ref{fig:approach}.
It requires the algorithm designers to have a model generator for the class of true processes the algorithm accepts.
The algorithm then generates repeatedly for a true process one or more event logs, and for each event log a set of samples.

We propose to use statistical tests to evaluate the two properties. 
Property P1 needs an analysis of the relation between the expected $\mathcal{P}^T$ and the observed $\mathcal{P}^S$.
For property P2, the Spearman rank correlation can be used to test whether there is a strong correlation between the sample quality and the log quality. 
If this is the case, then statistical evidence has been provided for the relationship between log and model quality.

\begin{table}[t]
	\centering
	\caption{Results of the controlled experiment. The last 10 columns show the Spearman rank correlation between the error measures, and precision and recall. All bold values are statistically significant ($p < 0.001$).}\label{tbl:run1}
	{\tiny
		\begin{tabular}{r|rr|rrrrr|rrrrr}
			& \multicolumn{2}{c|}{\textbf{True Process}} & \multicolumn{5}{c|}{\textbf{Precision}} & \multicolumn{5}{c}{\textbf{Recall}} \\\hline
			\textbf{Model} & \textbf{prec.} & \textbf{recall}        & Cov.           & sMAPE           & sRMSPE          & NRMSE           & NMAE            & Cov.              & sMAPE           & sRMSPE          & NRMSE             & NMAE            \\\hline
			1  & 0.538          & 1.000                  & \textbf{0.658} & \textbf{-0.988} & \textbf{-0.986} & \textbf{-0.988} & \textbf{-0.989} &  \textbf{0.338}   & \textbf{-0.356} & \textbf{-0.354} & \textbf{-0.354}   & \textbf{-0.356} \\
			2  & 0.797          & 1.000                  & \textbf{0.470} & \textbf{-0.986} & \textbf{-0.985} & \textbf{-0.901} & \textbf{-0.954} &  0.154            & -0.051          & -0.052          &  0.012            & -0.004          \\
			3  & 0.935          & 1.000                  & \textbf{0.781} & \textbf{-0.990} & \textbf{-0.989} & \textbf{-0.975} & \textbf{-0.984} &  \textbf{0.637}   & \textbf{-0.406} & \textbf{-0.417} & \textbf{-0.410}   & \textbf{-0.412} \\
			4  & 0.953          & 1.000                  & \textbf{0.705} & \textbf{-0.991} & \textbf{-0.992} & \textbf{-0.984} & \textbf{-0.987} & -0.103            &  0.105          &  0.108          &  0.081            &  0.090          \\
			5  & 0.988          & 1.000                  & \textbf{0.540} & \textbf{-0.983} & \textbf{-0.981} & \textbf{-0.980} & \textbf{-0.986} &  \textbf{0.437}   & -0.201          & -0.206          & -0.207            & -0.201          \\
			6  & 0.871          & 1.000                  & \textbf{0.532} & \textbf{-0.934} & \textbf{-0.938} & \textbf{-0.917} & \textbf{-0.926} & \textbf{-0.529}   &  \textbf{0.973} &  \textbf{0.962} &  \textbf{0.963}   &  \textbf{0.968} \\
			7  & 0.943          & 1.000                  & \textbf{0.511} & \textbf{-0.991} & \textbf{-0.989} & \textbf{-0.986} & \textbf{-0.989} &  \textbf{0.456}   & -0.242          & -0.240          & -0.228            & -0.231          \\
			8  & 0.616          & 1.000                  & \textbf{0.773} & \textbf{-0.992} & \textbf{-0.991} & \textbf{-0.989} & \textbf{-0.990} &  0.114            & -0.148          & -0.154          & -0.156            & -0.157 \\
			9  & 0.710          & 1.000                  & \textbf{0.519} & \textbf{-0.981} & \textbf{-0.978} & \textbf{-0.970} & \textbf{-0.973} &  \textbf{0.518}   & \textbf{-0.327} & \textbf{-0.330} & \textbf{-0.340}   & \textbf{-0.341} \\
			10 & 0.883          & 1.000                  & \textbf{0.703} & \textbf{-0.982} & \textbf{-0.982} & \textbf{-0.977} & \textbf{-0.976} &  0.116            & -0.022          & -0.027          & -0.016            & -0.023          \\
		\end{tabular}
	}
\end{table}

\begin{figure}[t]
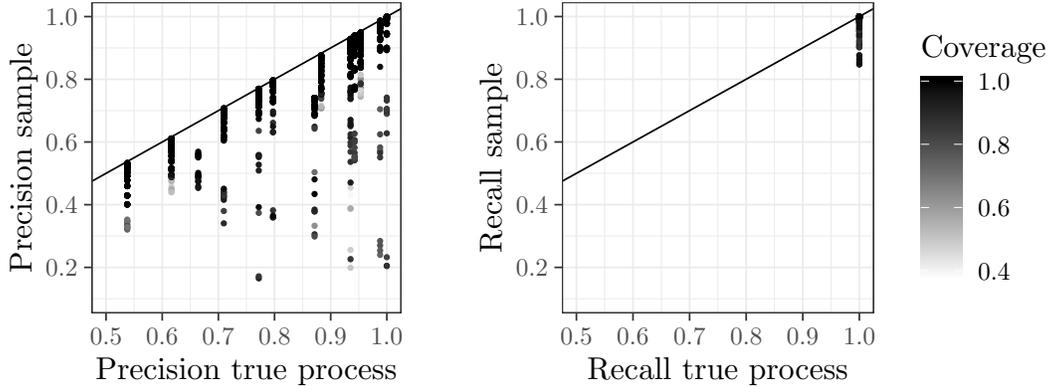

	\hfill
	\begin{minipage}{.385\textwidth}
		\resizebox{\columnwidth}{!}{\input{images/runtwo_precisionsample_precisiontrue_coverage.tex}}
	\end{minipage}
	\hfill
	\begin{minipage}{.555\textwidth}
		\resizebox{\columnwidth}{!}{\input{images/runtwo_recallsample_recalltrue_coverage.tex}}
	\end{minipage}
	\caption{Relation between the quality of the true process and the quality of the discovered models, for precision (left) and recall (right). Darker points represent a higher coverage.}\figlabel{im-propertyR1}
\end{figure}

\vspace{-2mm}
\subsubsection{Example Evaluation.}
\vspace{-1mm}
As an example, the controlled experiment has been implemented in ProM\footnote{The source code is available on: \url{https://github.com/ArchitectureMining/SamplingFramework}} for the Inductive Miner. 
To calculate precision and recall, an implementation of exact matching entropy-based measures in Entropia is used~\cite{PolyvyanyyACGKL20}.
For each true process, a single event log with 5,000 traces has been generated. 
The event logs were 10 times sampled for 12 sampling ratios: $0.01$, $0.02$, $0.05$, and $0.1$ up to $0.9$.

The results are shown in Tbl.~\ref{tbl:run1} and Fig.~\ref{fig:im-propertyR1}.
From Fig.~\ref{fig:im-propertyR1} we conclude that property P1 holds for precision and recall. 
For each model that describes the true process, the Spearman rank correlation is calculated between each of the log quality measures and precision, and similarly for recall.
As for the measures sMAPE, sRMSPE, NRMSE, and NMAE, $0$ is the best quality, a negative correlation indicates the required guarantee that samples of higher quality results in better discovered models, whereas for coverage, a positive correlation indicates this result. 
As can be seen in the table, the experiment generates mixed results. 
Though property P2 holds for precision, it is not satisfied for recall.
Hence, we can conclude that the Inductive Miner satisfies the two properties for precision, but fails to do so for recall on the second property.

\subsection{Stage 4: The Algorithm is Effective}
An established relationship between log and model quality, the essence of stage 3, does not guarantee the algorithm to be effective in real-life situations.
The main caveat in the controlled environment of the previous stage is that the true process is known.
Each event log is generated from the known true processes.
In real-life situations, the true process is unknown, and, hence, may invalidate assumptions of the discovery algorithm.
For example, the Inductive Miner assumes event logs to be generated from process trees.
However, no criteria are given to test whether an event log is generated by a process tree, 
nor does the algorithm provide any details on the model quality if the assumption is invalid.

In this stage, the algorithm designer has to validate how effective the algorithm is in real-life situations.
One way to obtain insights into the effectiveness of the algorithm is to apply sampling on a benchmark.
This benchmark can be a set of well-known real-life event logs as used in~\cite{AugustoCDRMMMS19}, or can be generated automatically, if the designers ensure that the class of generated models is larger than the class of true processes studied in the previous stage.
The algorithm designers need to analyze property P2 in the absence of a true process.
In other words, even if the true process is unknown, event logs of better quality should return better quality models. 
This may result in an experiment as outlined in Algorithm~\ref{algo:invivo}.

The analysis of property P2 in the absence of a true process can have two possible outcomes.
Either it is shown that the algorithm has the desired property, or, if this is not possible, the algorithm should be further improved, or provide additional log quality measures, that guarantee that an event log satisfies the assumptions of the process discovery algorithm.

\begin{figure}
	\begin{minipage}{\textwidth}
		\begin{minipage}{.46\textwidth}
			\resizebox{\columnwidth}{!}{\input{images/road_errors_ratio.tex}}
		\end{minipage}
		\hfill
		\begin{minipage}{.46\textwidth}
			\resizebox{\columnwidth}{!}{\input{images/sepsis_errors_ratio.tex}}
		\end{minipage}
		\vspace{-4mm}
		\caption{Plot of ratio and the sample quality measures coverage ($\bullet$), sMAPE ($+$), sRMSPE ($\boxtimes$), NRMSE ($\blacksquare$) and NMAE ($\blacktriangle$) for the Road Fine log (left) and the Sepsis log (right).}\label{fig:sepsis-ratio-error}\label{fig:road-ratio-error}
	\end{minipage}
	\begin{minipage}{\textwidth}
		\begin{minipage}{.46\textwidth}{\scriptsize
			\resizebox{\columnwidth}{!}{\input{images/road_precision_ratio.tex}}
			\vspace{-2mm}\resizebox{\columnwidth}{!}{\input{images/road_precision_coverage.tex}}
			\vspace{-2mm}\resizebox{\columnwidth}{!}{\input{images/road_recall_coverage.tex}} }
		\end{minipage}
		\hfill
		\begin{minipage}{.46\textwidth}{\scriptsize
			\resizebox{\columnwidth}{!}{\input{images/sepsis_precision_ratio.tex}}
			\vspace{-2mm}\resizebox{\columnwidth}{!}{\input{images/sepsis_precision_coverage.tex}}
			\vspace{-2mm}\resizebox{\columnwidth}{!}{\input{images/sepsis_recall_coverage.tex}} }
		\end{minipage}
		\vspace{-5mm}
		\caption{Plots of ratio and precision, and coverage with precision and recall for the Road Fine log (left) and the Sepsis log (right).}\label{fig:road-error-precision}\label{fig:sepsis-error-precision}\label{fig:sepsis-ratio-precision}\label{fig:road-ratio-precision}
	\end{minipage}
\end{figure}

\subsubsection{Example Evaluation.}
As an example of an analysis in stage 4, we conducted the proposed experiment on the Inductive Miner. 
Two real-life event logs have been selected, the Road Traffic Fine management process event log~\cite{deLeoni2015} and the Sepsis cases event log~\cite{Mannhardt2016}.
The Road Fine log has in total 150,370 traces and 561,470 events.
There are 231 unique traces and 11 unique event types.
The Sepsis log consists of 1,049 traces, of which 845 are unique, and 15,190 events with 16 unique event types.
Sampling was done at the same sampling ratios as before: $0.01$, $0.02$, $0.05$, and $0.1$ up to $0.9$.
For each ratio, 10 samples were drawn.

The sample quality measures for the Road Fine log are shown on the left in Fig.~\ref{fig:road-ratio-error}.
As the plot shows, the larger the sampling ratio, and thus the log size, the better the quality is (error measures: $\rho < -0.9$, $p < 0.001$, coverage: $\rho=0.96$, $p<0.001$).
Sample size and the conformance measure on precision (Fig.~\ref{fig:road-ratio-precision}) shows a moderate positive correlation ($\rho=0.56$, $p < 0.001$), while there is no correlation between sampling ratio and recall ($\rho=0.03$, $p=0.72$).
Analyzing the quality measures with the conformance measures shows a different story.
In Fig.~\ref{fig:road-error-precision}, the coverage is plotted against the precision, indicating there is no correlation between coverage and precision.
Further analysis found there are no correlations between the sample quality measures and precision 
(sMAPE: $\rho = -0.19$, $p=0.03$, sRMSPE: $\rho = -0.18$, $p=0.051$, NRMSE: $\rho = -0.21$, $p=0.02$, NMAE: $\rho = -0.20$, $p=0.03$, coverage: $\rho=0.17$, $p=0.06$).
The correlations found for recall show that samples of worse quality result in better models
(sMAPE: $\rho = 0.80$, $p<0.001$, sRMSPE: $\rho = 0.79$, $p< 0.001$, NRMSE: $\rho = 0.77$, $p<0.001$, NMAE: $\rho = 0.78$, $p < 0.001$, coverage: $\rho=-0.79$, $p<0.001$).

For the Sepsis log, similar results are found.
As indicated by the plots at the right hand side of Fig.~\ref{fig:sepsis-ratio-error}, a correlation is found between the sampling ratio and the log quality measures (for all error measures: $\rho < -0.9$, $p < 0.001$, coverage: $\rho=0.59$, $p<0.001$).
The larger the sampling ratio, the higher the precision is ($\rho=0.57$, $p< 0.001$), but
no correlation was found between sampling ratio and recall ($\rho=0.03$, $p=0.72$).
A moderate negative correlation was found between the log quality measures and precision
(for the error measures: $-0.60 < \rho < -0.50$, $p < 0.001$, coverage: $\rho=0.59$, $p<0.001$),
while the log quality measures did not show any correlation with recall 
(for all measures: $-0.04 < \rho < 0.02$, $p>0.70$).

\smallskip
As the results show, there is no clear relation between log and model quality.
Hence, it is with the current measures not possible to conclude that the Inductive Miner is guaranteed to be effective in real-life situations.
As a next step, new log quality measures should be developed that do establish the required relationship between log and model quality. 
The process can then be repeated until sufficient guarantees can be provided on the effectiveness of the algorithm.

\section{Related Work}\label{sec:relatedwork}
The statistical approach we propose to establish a relation between log and model quality relates to event data quality in general, builds upon established properties of conformance measures, and requires sampling techniques on event logs. 
In this section, we review literature on these topics, and show how our approach relates to these.

\paragraph{Measuring log quality.}
As the process mining manifesto articulates, process mining treats data as first-class citizens~\cite{manifesto}, and defines four data qualities,
of which \emph{completeness} is studied mostly.
For example, \cite{BoseMA2013} identifies four categories of process characteristics and 27 classes of event log quality issues. 
Most studies in event log quality focus on the incompleteness of the data. 
Examples include not having enough information recorded in the event log (e.g. missing cases or events)~\cite{BoseMA2013,Aalst2016}, not having recorded enough behavior in the event log~\cite{Gunther2009}, or the traces not being representative of the process~\cite{Gunther2009}, and noise.
Different notions of noise are studied, such as infrequent behavior that is either incorrect or rare~\cite{MedeirosWA07}.
However, event logs are studied in isolation in these studies. 
Instead, we argue to assess the quality of event logs relative to other event logs, using statistical techniques based on sampling.

\paragraph{Properties of conformance measures.}
The process mining community has recently initiated a discussion on which formal properties should ``good'' conformance measures satisfy.
In~\cite{TaxLSFA18}, the authors proposed five properties for precision measures.
For instance, one property states that for two process models that describe all the traces in the log, a less permissive model should not be qualified as less precise.
By demonstrating that a measure fulfills such properties, one establishes its usefulness.
In~\cite{PolyvyanyySWCM20}, the authors strengthened the properties from~\cite{TaxLSFA18}.
For example, according to these properties, the less permissive model from the example above should be classified as more precise.
In~\cite{Aalst2018ataed}, the precision properties from~\cite{TaxLSFA18} were refined, and further desired properties for recall and generalization measures were introduced, resulting in 21 conformance propositions.
Finally, in~\cite{PolyvyanyyK19}, properties for precision and recall measures that account for the partial matching of traces, i.e., traces that are not the same but share some subsequences of activities, were introduced.
The precision and recall measures used in our evaluations satisfy all the introduced desired properties for the corresponding measures~\cite{TaxLSFA18,PolyvyanyySWCM20,Aalst2018ataed,Syring2019}.

\paragraph{Sampling in process mining.}
Sampling has been studied before in process mining, but never as a systematic approach to evaluate process discovery techniques.
A first set of measures for the representativeness of samples have been proposed in~\cite{KnolsW19}. 
Their results show the need for a systematic approach as proposed in this paper.

In~\cite{Berti2017}, a sampling technique specific for the Heuristics Miner is described, claiming that only 3\% of the original log is sufficient to discover 95\% of the dependency relations.
However, a proper evaluation of this claim has not been provided, nor are the results generalizable to other process discovery techniques.

A statistical framework based on \emph{information saturation} is proposed in~\cite{Bauer2018}.
Their approach differs from the probability sampling techniques we propose. 
Instead of generating samples that estimate the event log, their approach focuses on creating a sufficiently small sample that contains as much information from the event log as possible. 
Consequently, this approach cannot be used to measure sample quality with respect to the event log.

A set of four biased sampling techniques is described in~\cite{SaniZA20}.
These techniques have been evaluated on six real-life event logs and three discovery techniques. 
The evaluation showed that sampling sometimes improves the F-measure for some of the models. 
A similar result on the F-measure was obtained in~\cite{LiuPZD2018}. Their study applied the Google PageRank algorithm on event logs to create a representative sample, which reduced the execution time of the Inductive Miner by half, without decreasing the F-measure. 
As the F-measure harmonizes precision and fitness, and no analysis was performed on the reasons behind the improvements, it is unclear how sampling influenced the process discovery results of both studies.
Instead of using sampling to improve the quality of the output, we propose to use probability sampling to analyze the input of algorithms, and to establish a relationship between log and model quality.
This relationship then allows to analyse why some samples give better models than other samples.

\section{Conclusion}\label{sec:conclusion}
This paper identifies the need for process discovery algorithms with guarantees that characterize the dependency between the quality of input event logs and the quality of the process models constructed from these event logs.
In particular, we argue that process discovery algorithms should produce better models from better input logs.
Currently, process discovery algorithms have never provided such guaranties, since, so far, we as a community, lacked a theoretical foundation to establish such a relationship.
In this paper, for the first time, measures for the statistical sample quality for ranking the quality of event logs are proposed.
We recommend using standard conformance checking measures for assessing the quality of the discovered models.
Combining log quality measures with conformance measures provides a framework to formally define properties that express the desired guarantee that better event logs result in better models.
These properties can be instantiated with various measures for quality of event logs and process models and be less or more pronounced, for example, imposing a strictly increasing or non-decreasing relation, or requiring a statistical association of a certain degree between the qualities of the corresponding logs and models.
To overcome this problem, we propose four stages in the design of an algorithm. 
Each design comes with additional properties and obligations to establish effective algorithms with guarantees.

We invite the process mining community to further contribute to the discussion of desired qualities for process discovery algorithms to ensure that state-of-the-art algorithms fulfill them.

\smallskip
\noindent
\textbf{Acknowledgments.}
Artem Polyvyanyy was in part supported by the Australian Research Council project DP180102839.

\bibliographystyle{plain}
\bibliography{bibliography}
\end{document}